\documentclass[twocolumn,secnumarabic,amssymb, nobibnotes, superscriptaddress, nofootinbib,aps,prd]{revtex4-2}
\usepackage[colorlinks,citecolor=blue,linkcolor=blue,anchorcolor=blue,filecolor=blue, urlcolor=blue]{hyperref}
\usepackage{natbib}

\usepackage{amsmath}
\usepackage{adjustbox}
\usepackage{graphicx}
\usepackage{booktabs}       
\usepackage{amsfonts}       
\usepackage{nicefrac}       
\usepackage{microtype}
\usepackage{cleveref}
\usepackage{xcolor}
\usepackage{multirow}
\usepackage{makecell}
\usepackage{array}
\usepackage{orcidlink}
\usepackage{float}

\begin{document}

\author{Adamu Issifu \orcidlink{0000-0002-2843-835X}} 
\email{ai@academico.ufpb.br}
\affiliation{Departamento de F\'isica, Instituto Tecnol\'ogico de Aeron\'autica, DCTA, 12228-900, S\~ao Jos\'e dos Campos, SP, Brazil} 
\affiliation{Laborat\'orio de Computa\c c\~ao Cient\'ifica Avan\c cada e Modelamento (Lab-CCAM)}

\author{Andreas Konstantinou~\orcidlink{0000-0002-1072-7313}}\email{akonst29@ucy.ac.cy}
\affiliation{Department of Physics, University of Cyprus, P.O. Box 20537, 1678 Nicosia, Cyprus}

\author{Prashant Thakur~\orcidlink{0000-0001-5930-7179}}
\email{prashantthakur1921gmail.com}
\affiliation{Department of Physics, Yonsei University, Seoul, 03722, South Korea}

\author{Tobias Frederico \orcidlink{0000-0002-5497-5490}} 
\email{tobias@ita.br}

\affiliation{Departamento de F\'isica, Instituto Tecnol\'ogico de Aeron\'autica, DCTA, 12228-900, S\~ao Jos\'e dos Campos, SP, Brazil} 
\affiliation{Laborat\'orio de Computa\c c\~ao Cient\'ifica Avan\c cada e Modelamento (Lab-CCAM)}

\title{Rotating PNS Admixed with Mirror Dark Matter: A Two-Fluid Approach} 
\begin{abstract}
  This work investigates the impact of mirror dark matter (DM) on the global properties of rotating neutron stars (NSs) across evolutionary stages, from hot, lepton-rich protoneutron stars (PNSs) to cold, catalyzed NSs along the Kelvin–Helmholtz timescale. The baryonic matter (BM) is modeled using a relativistic mean-field (RMF) approach with density-dependent couplings, while the dark sector mirrors the visible sector with analogous thermodynamic conditions. Using a two-fluid formalism with purely gravitational DM-BM interaction, we find that rotation enlarges the star, whereas DM admixture increases compactness and enhances gravitational stability. However, increased compactness due to DM lowers the threshold for rotational instabilities, making DM-admixed stars more susceptible. Rotation decreases {central temperature behavior} by redistributing thermal energy over a larger volume and reducing central density, while DM raises temperatures by deepening the gravitational potential and increasing thermal energy. Stars become more prone to collapse and rotational instabilities as frequency ($\nu$) rises and the polar-to-equatorial radius ratio ($r_p/r_e$) decreases, especially near the Keplerian limit ($\nu_K$). DM-admixed stars also show higher surface gravitational redshifts due to their compactness. Our results qualitatively agree with universal relations primarily derived for rotating cold stars. These findings highlight competing effects of rotation and DM on NS thermal evolution, structure, and observables, potentially offering indirect probes of DM within NSs.
  
\end{abstract}

\maketitle

\section{Introduction}

{Observational evidence from large-scale cosmic structures, gravitational lensing, and galactic rotation curves supports the existence of DM. According to standard cosmological models, approximately 94\% of the universe’s total mass-energy content consists of DM and dark energy, while ordinary (baryonic) matter accounts for only about 6\% \cite{Massey:2010hh, Lelli:2016zqa, Planck:2018vyg}. Despite extensive experimental efforts, direct detection of DM in terrestrial laboratories remains elusive, primarily due to limited knowledge of its fundamental properties, interaction channels, mass, charge, and its coupling to BM \cite{PandaX-II:2017hlx, CRESST:2019jnq, XENON:2018voc}. Given its well-established astrophysical signatures, many theoretically motivated models treat DM as a fundamental particle to investigate its possible properties. However, despite decades of direct and indirect searches, both in laboratory experiments and through astrophysical modeling and observations, its true nature remains unknown, driving ongoing investment and intensive research aimed at detection and characterization~\cite{Klasen:2015uma, Schumann:2019eaa, Liu:2017drf}.

There are two main approaches to investigating the presence of DM in NSs, motivated by the complexity of DM interactions and the limited understanding of its accumulation and interaction mechanisms within NSs. These approaches are commonly categorized as two-fluid and single-fluid formalisms. In the two-fluid formalism, DM is assumed to interact with BM exclusively through gravity, allowing both components to coexist but evolve independently in the star's structure \cite{Kain:2021hpk, Leung:2011zz, Rutherford:2022xeb, Issifu:2024htq, Thakur:2024btu, Shakeri:2022dwg, Miao:2022rqj, Buras-Stubbs:2024don}. The stellar structure in this case is determined by solving the two-fluid Tolman-Oppenheimer-Volkoff (TOV) equations, treating DM and BM as two distinct interpenetrating fluids. In contrast, the single-fluid formalism assumes that DM interacts with BM via Standard Model particles or other mediators, resulting in a coupled system described by a single equation of state \cite{Lenzi:2022ypb, Lourenco:2022fmf, Das:2018frc, Lopes:2024ixl}. The star’s structure in this scenario is modeled using the standard single-fluid TOV equations. 

Given the observational advantages offered by NSs over Earth-based experiments, the presence of DM is expected to affect several observable NS properties, including their mass, radius, and tidal deformability \cite{Leung:2022wcf}. Additionally, DM can influence internal dynamical properties such as rotational behavior, angular momentum loss, and even gravitational wave emission, due to potential modifications in the star’s photon signatures and oscillation modes \cite{Rutherford:2022xeb, Nelson:2018xtr, Bramante:2013hn}. However, the extent to which DM accumulates and alters the stellar structure strongly depends on key factors such as the DM’s initial velocity distribution, its scattering cross-section with BM, the gravitational potential of the star, and its escape velocity. For example, if the star captures self-annihilating DM \cite{Kouvaris:2007ay, Bertone:2007ae}, such as the weakly interacting massive particles (WIMPs) \cite{Arcadi:2017kky, LUX:2015abn}, the resulting annihilation processes can heat the stellar interior, thereby affecting its thermal evolution history. In contrast, the accumulation of non-self-annihilating DM -- such as asymmetric DM \cite{Petraki:2013wwa, Zurek:2013wia} or mirror DM \cite{Issifu:2024htq, Kouvaris:2010jy} -- primarily alters the star’s macroscopic structure, impacting its tidal deformability, oscillation frequencies, and gravitational wave signatures. These distinct effects offer promising observational pathways for constraining DM properties using astrophysical data.

This study investigates how mirror DM influences the global structural evolution of PNSs, from their initial neutrino-rich phase to their final state as cold, neutrino-transparent compact stars. PNSs are formed following the core collapse of massive stars that have exhausted their nuclear fuel, leading to a supernova explosion \cite{Janka:2012wk, Kotake:2005zn}. The analysis covers the star’s evolution from the early phase, when it is lepton-rich and neutrino-trapped, through the deleptonization stage, characterized by neutrino diffusion out of the stellar core following core bounce, to the neutrino-transparent phase, and finally to the formation of a cold, catalyzed NS \cite{Pons:1998mm, Prakash:1996xs, Issifu:2024fuw}. The equation of state (EoS) for BM and DM is computed separately, with their interaction introduced purely through gravity via the two-fluid formalism. The BM sector is modeled using the RMF approach with density-dependent couplings \cite{Menezes:2021jmw, Roca-Maza:2011alv}. For the dark sector, we adopt a mirror matter framework, assuming that dark protons and dark neutrons interact through a mirrored version of the strong force, analogous to their visible counterparts. The details of the two models can be found in \cite{Issifu:2024htq}.  

For this study, we adopt and modify the standard \texttt{rns} codes, originally developed for modeling rotating NSs in a single-fluid framework, to model a two-fluid system consisting of BM and DM (DM-admixed NSs or DMANSs) \citep{Konstantinou:2024ynd, Stergioulas:1994ea}. For simplicity, we assume that the DM component remains static and non-rotating within the stellar core, while the BM fluid rotates around it. This approach allows us to isolate the gravitational effects of DM on the rotating stellar structure. This approach is motivated by the assumption of weak interaction between the two matter components, where DM interacts with BM solely through gravity \cite{Sandin:2008db}. Under this assumption, the DM fluid contributes negligible viscosity and carries no angular momentum, allowing us to treat the DM as non-rotating while all the angular momentum resides in the BM sector. This assumption helps isolate and better understand the effects of DM on key observables such as mass, radius, rotational deformation, and moment of inertia \cite{Konstantinou:2024ynd, Cronin:2023xzc}. This simplification realistically mimics a DMANS scenario where DM forms a static gravitational well inside the star. It also provides clearer physical insight into how DM influences the BM’s density distribution, stability, and rotational dynamics, and significantly reduces computational costs. 

For the first time, we apply this approach to investigate the impact of DM on the structural evolution of PNSs. Our focus is on how rotational frequency influences key global properties, such as mass, radius, and moment of inertia, as the PNS evolves through its different post-birth stages. Additionally, we explore how thermodynamic conditions, including lepton fraction and entropy per baryon, affect DM accumulation during stellar rotation and evolution. Special attention is given to the role of neutrino pressure, particularly during the early evolutionary phases, in shaping the rotational dynamics. The influence of DM is then studied by treating it as a gravitationally-interacting fluid with pre-determined mass fractions, allowing us to isolate its structural effects.

The paper is organized as follows. In \cref{md}, we describe our model, covering the EoS for the BM, thermodynamic constraints, the EoS for DM, the rotating two-fluid formalism for DMANSs, and the structural universality relations for DMANSs, discussed in subsections \cref{bm}, \cref{tc}, \cref{dm}, \cref{sec:2fluid_summary}, and \cref{sec:universal_mass_radius}, respectively. Our results and detailed discussion are presented in \cref{rs}. Finally, we provide our concluding remarks in \cref{conc}.
}
\section{Model Description and Formalism} \label{md}
\subsection{The EoS of the BM} \label{bm}

\begin{table}[t]
\caption {DDME2 parameters.}
\begin{center}
\begin{tabular}{ |c| c| c| c| c| c| c| }
\hline
 meson($i$) & $m_i(\text{MeV})$ & $a_i$ & $b_i$ & $c_i$ & $d_i$ & $g_{i N} (n_0)$\\
 \hline
 $\sigma$ & 550.1238 & 1.3881 & 1.0943 & 1.7057 & 0.4421 & 10.5396 \\  
 $\omega$ & 783 & 1.3892 & 0.9240 & 1.4062 & 0.4775 & 13.0189  \\
 $\rho$ & 763 & 0.5647 & $\cdots$ & $\cdots$ & $\cdots$ & 7.3672 \\
 \hline
\end{tabular}
\label{T}
\end{center}
\end{table}

{The EoS of the BM is derived within the quantum hadrodynamics framework, typically modeled via the RMF theory, where hadronic interactions are mediated by the exchange of massive scalar and vector mesons (see a review in Ref.~\cite{Menezes:2021jmw}). We adopt a model governed by the Lagrangian density given by 
\begin{align}
 \mathcal{L}_{\rm H}{}&=  \sum_{N=p,n} \bar \psi_N \Big[  i \gamma^\mu\partial_\mu - \gamma^0  \big(g_{\omega N} \omega_0  +  g_{\rho N} I_{3} \rho_{03}  \big)\nonumber\\
 &- \Big( m_N- g_{\sigma N} \sigma_0 \Big)  \Big] \psi_N, \label{h}\\
\mathcal{L}_{\rm m}&= - \frac{1}{2} m_\sigma^2 \sigma_0^2  +\frac{1}{2} m_\omega^2 \omega_0^2   +\frac{1}{2} m_\rho^2 \rho_{03}^2,\label{m}\\
 \mathcal{L}_{\rm l}& = \sum_l\Bar{\psi}_l\left(i\gamma^\mu\partial_\mu-m_l\right)\psi_l\label{l},
\end{align}
where $\psi_N$ denotes the Dirac field for baryons (protons and neutrons), and $\sigma_0$, $\omega_0$, and $\rho_{03}$ correspond to the mean-field representations of the scalar meson, vector-isoscalar meson, and the third component of the vector-isovector meson, respectively. Their masses are denoted by $m_i$ ($i = \sigma,\, \omega,\,\rho$). The nucleon-meson coupling constants are represented by $g_{iN}$, $I_3 = \pm 1/2$ denotes the isospin projection of the nucleons, and $m_N=938$MeV is the nucleon mass. $\psi_l$ represents the free lepton field, introduced to maintain charge neutrality in the stellar matter and to ensure the correct lepton fraction at each stage of PNS evolution, in accordance with supernova remnant physics~\cite{Issifu:2023qyi, Issifu:2023qoo, Prakash:1996xs, Pons:1998mm}. The meson-nucleon couplings are calibrated by the DDME2 parameterization given by 
\begin{equation}
    g_{i N} (n_B) = g_{iN} (n_0)a_i  \frac{1+b_i (\eta + d_i)^2}{1 +c_i (\eta + d_i)^2},
\end{equation}
and 
\begin{equation}
    g_{\rho N} (n_B) = g_{\rho N} (n_0) \exp\left[ - a_\rho \big( \eta -1 \big) \right].
\end{equation}
Here, $\eta = n_B/n_0$, where $n_B$ denotes the baryon density and $n_0 = 0.152\,\text{fm}^{-3}$ represents the nuclear saturation density within this model framework. The model parameters $a_i$, $b_i$, $c_i$, and $d_i$ are calibrated to reproduce bulk nuclear matter properties around the saturation density $n_0$. The complete set of parameters, along with the meson masses, is presented in \Cref{T}.

\subsubsection{Thermodynamic Conditions for the PNS Evolution}\label{tc}
The corresponding EoS for the stellar matter is determined by solving the energy-momentum conservation equations derived from the Lagrangian density
$$
\mathcal{L}_{\rm RMF} = \mathcal{L}_{\rm H} + \mathcal{L}_{\rm m} + \mathcal{L}_{\rm l},
 $$
imposing conditions of $\beta$-equilibrium, charge neutrality, and baryon number conservation. A full derivation of the EoS can be found in Refs.~\cite{Issifu:2023qyi, Issifu:2024fuw, Issifu:2024htq} and references therein. To avoid repetition, we present only the relevant highlights here. The thermal effects and the EoS can be determined collectively from the Helmholtz free energy relation

 $$
 \mathcal{F} = \varepsilon_{\rm BM} - s\,T,
 $$
 where $\varepsilon_{\rm BM}$ is the total energy density, $T$ is the temperature, and $s$ is the entropy density. The main equation for calculating the $\beta$-equilibrated EoS, which connects $\varepsilon_{\rm BM}$, the total pressure $P_{\rm BM}$, $T$, the chemical potentials $\mu_j$ of the individual particles, $j$, present, and the $n_B$ for the neutrino-trapped matter phase, is given by
\begin{equation}
    sT=P_{\rm BM}+\varepsilon_{\rm BM} -n_B\mu_B-\mu_{\nu_e}(n_{\nu_e} +n_e),
\end{equation}
where $\mu_{\nu_e}$ and $\mu_B$ are the neutrino and the baryon chemical potentials respectively. $n_B$, $n_e$, and $n_{\nu_e}$ are the baryon, electron, and neutrino number densities, respectively. The neutrino-transparent phase is determined from 
\begin{equation}
    sT=P_{\rm BM}+\varepsilon_{\rm BM} -n_B\mu_B.
\end{equation}
To track stellar evolution, we divide our analysis into two phases:
\begin{enumerate}
    \item \textbf{Neutrino-trapped phase:} Here, we model the neutrino-trapping phase of the PNS by fixing the lepton fraction at $Y_l = 0.4$ and the entropy per baryon at $s_B = 1$. For the subsequent deleptonization phase after core bounce, we set $Y_l = 0.2$ and $s_B = 2$.
    
    \item \textbf{Neutrino-transparent phase:} At this stage, we model the star after all neutrinos have escaped from the stellar core, and the temperature of the stellar matter has reached its peak and begins to cool. Finally, we model the cold and catalyzed NS configuration corresponding to this phase.
\end{enumerate}
The approach adopted for modeling the evolution of the PNSs follows the approach detailed in Ref.~\cite{Pons:1998mm, Prakash:1996xs, Vartanyan:2018iah, Camelio:2017nka}.

\subsection{The EoS of the DM}\label{dm}
We adopt a mirror DM model that replicates the BM sector in terms of interaction and coupling constraints similar to the one presented in \cite{Issifu:2024htq}:
\begin{align}\label{L1}
    {\cal L}_{\rm DM}={}&{\overline\psi_D}[ (i\gamma_\mu\partial^\mu- \gamma^0g_{v} V_0)-(m_D -g_{\tilde{\sigma}} \tilde{\sigma}_0)]\psi_D \nonumber\\
    -&\frac{1}{2}m_{\rm \tilde{\sigma}}^{2}\tilde{\sigma}_0^{2}
    +\frac{1}{2}m_{v}^{2} V_0^2.
\end{align}
Here, $V_0$ (with mass $m_v$ and coupling $g_v$) and $\tilde{\sigma}$ (with mass $m_{\tilde{\sigma}}$ and coupling $g_{\tilde{\sigma}}$) correspond to the $\omega_0$ and $\sigma_0$ mesons in the visible sector, representing analogous fields and properties in the dark sector. Also, $\psi_D(\overline\psi_D)$ represents the mirror DM fields. However, in this sector, we treat dark protons and dark neutrons as identical particles under the strong interaction; consequently, the isospin projection is omitted from Eq.~(\ref{L1}). The advantage of this approach is that dark PNSs can be modeled in isolation, with their stellar matter satisfying dark $\beta$-equilibrium, dark charge neutrality, and dark baryon number conservation. This enables the use of a DM EoS with the same thermodynamic properties as in the visible sector, without the need to account for thermal equilibrium conditions between the two matter components, like the approach discussed in Ref.~\cite{Issifu:2024htq}. A similar model has been used in Refs.~\cite{Xiang:2013xwa, Thakur:2023aqm, Marzola:2024ame} to study DMANSs, where the DM is treated as a neutral field. The relevant equations for calculating the EoS are as follows. For neutrino-transparent matter:
\begin{equation}
\tilde{s}T_D = P_D + \varepsilon_D - n_{B'} \mu_{B'},
\end{equation}
where $T_D$ is the DM temperature, $P_D$ the DM pressure, $\varepsilon_D$ the DM energy density, $\tilde{s}$ the DM entropy density, $n_{B'}$ the dark baryon density, and $\mu_{B'}$ the dark chemical potential. For neutrino-trapped matter, the equation becomes:
\begin{equation}
\tilde{s} T_D = P_D + \varepsilon_D - n_{B'} \mu_{B'} - \mu_{\nu_e'} \left( n_{\nu\_e'} + n_{e'} \right),
\end{equation}
where $n_{e'}$ is the dark electron number density, $n_{\nu_e'}$ the dark neutrino number density, and $\mu_{\nu_e'}$ the dark neutrino chemical potential.

}
\subsection{Rotating Two-Fluid DMANS Stars}
\label{sec:2fluid_summary}

To describe the structure of rotating DMANS, we use a modified version of the \texttt{rns} code \citep{Konstantinou:2024ynd, Stergioulas:1994ea} where a stationary and axisymmetric spacetime is assumed. The metric employed is
\begin{equation}
\begin{aligned}
  ds^2 = -e^{\tilde{\gamma} + \tilde{\rho}} dt^2 + e^{2\tilde{\alpha}} (dr^2 + r^2 d\theta^2) \\
  + e^{\tilde{\gamma} - \tilde{\rho}} r^2 \sin^2\theta (d\phi - \tilde{\omega} dt)^2,
\end{aligned}
\label{eq:metric}
\end{equation}
where \( \tilde{\gamma}, \tilde{\rho}, \tilde{\alpha}, \tilde{\omega} \) are metric potentials that depend on the radial coordinate \( r \) and the polar angle \( \theta \).

If the NS accreted DM from its surroundings, a more realistic scenario would involve a torque-free differential rotation of the DM component induced by frame-dragging effects \cite{shawqi2025rotatingneutronstarsdark}. Moreover, the characteristic timescale for DM capture and accumulation ($10^6$–$10^9$ years) \cite{McKeen:2023ztq, Grippa:2024ach} is many orders of magnitude longer than the short dynamical timescale of PNS evolution (tens of seconds) \cite{Prakash:1996xs, Janka:2012wk, Pons:1998mm}. This motivates treating the DM as a pre-existing, static core during the short PNS phase. In this study, however, we assume that the DM fluid is non-rotating, as this work predates \cite{shawqi2025rotatingneutronstarsdark}, and therefore employs an earlier version of the RNS two-fluid code \cite{Konstantinou:2024ynd}, suitable for PNS modeling. Previous studies \cite{Konstantinou:2024ynd} suggest that even when the DM co-rotates with the BM, its impact on the overall stellar structure remains negligible, though future work will explore these dynamic coupling effects more comprehensively. 

Also, we assume that the two fluids are only interacting through gravity, and the BM is rotating with uniform angular velocity. Hence, the total energy-momentum tensor, $T_{tot}^{\mu \nu}$, can be written as a linear combination of the dark matter, $T_D^{\mu \nu}$, and the baryonic matter, $T_B^{\mu \nu}$, energy-momentum tensors
\begin{equation}
    T_{tot}^{\mu \nu}=T_D^{\mu \nu}+T_B^{\mu \nu}.
    \label{EN_MOM}
\end{equation}

The physical circumferential radii are defined as
\begin{align}
  R_{Be} &= r_{Be} e^{(\tilde{\gamma}_{Be} - \tilde{\rho}_{Be})/2}, \\
  R_{De} &= r_{De} e^{(\tilde{\gamma}_{De} - \tilde{\rho}_{De})/2}, \\
  R_{Bp} &= r_{Bp} e^{(\tilde{\gamma}_{Bp} - \tilde{\rho}_{Bp})/2},
\end{align}
with subscripts \( e \) and \( p \) denoting equatorial and polar values, respectively.

The total gravitational mass is given by
\begin{equation}\label{Tmass}
  M_{\text{tot}} = M_{GB} + M_{GD},
\end{equation}\label{tmass}
where \( M_{GB} \) and \( M_{GD} \) are the gravitational masses of baryonic and dark matter components, defined by equations A26 and A27 of \citep{Konstantinou:2024ynd}.
\\The redshift on the poles, $Z_p$, is determined as
\begin{equation}
  Z_p=e^{-(\tilde{\gamma}_{Bp}+\tilde{\rho}_{Bp})/2}-1.
\end{equation}

In what follows, when we refer to the dark matter mass fraction in this paper, we mean that
\begin{equation}
  f_\chi = \frac{M_{GD}}{M_{GD} + M_{GB}},
\end{equation}
is evaluated at the non-rotating limit, on a star with the same total central energy density, $\epsilon_c^{tot}=\epsilon_c^{BM}+\epsilon_c^{DM}$, and pressure, $P_c^{tot}=P_c^{BM}+P_c^{DM}$.

To run \texttt{rns}, one specifies the ratio of equatorial to polar radii, $r_\text{ratio} = \frac{r_e}{r_p}$.
For two-fluid stars, separate ratios are defined {
$ r_{\text{ratioB}} = \frac{r_{Be}}{r_{Bp}}$ for BM and 
  $r_{\text{ratioD}} = \frac{r_{De}}{r_{Dp}}$ for DM.}
As in this work we are assuming minimal spin for the dark matter component, then \( r_{\text{ratioD}} \approx 1 \).

The baryonic moment of inertia is computed using the formula \begin{equation}
  I_{BM} = 2\frac{T_{kin}}{ \Omega^2} \, ,
\end{equation}
where $\Omega=2\pi\nu$, with $\nu$ being the rotation frequency, and $T_{kin}$ is the kinetic energy. Notice that as the dark matter fluid is not rotating, its kinetic energy must be zero, and therefore, using here $T_{kin}$ or $T^{BM}_{kin}$, it does not make any difference.

Finally, the baryonic mass is defined as 
\begin{equation}
M_{b}=\int e^{2\overline{\alpha}+(\overline{\gamma}-\overline{\rho})/2}\frac{m_N n_B(r)}{(1-\upsilon^2)^{1/2}}\sin(\theta) r^2 drd\theta d\phi.
\label{eq:38}
\end{equation}
where $m_N$ and $n_B$ are the nucleon mass and baryon density, respectively, and $\upsilon=(\Omega-\overline{\omega})r\sin(\theta) e^{-\overline{\rho}}$. The conserved baryon mass, $M_b$, is obtained by integrating the baryon number density over the curved spacetime geometry, which naturally includes the total gravitational influence of both the baryonic and DM components. This formulation captures the effect of the DM's gravitational field on the baryonic mass distribution while preserving the fact that DM itself carries no baryon number. Consequently, this expression ensures that both the baryon number of OM and the DM particle number are separately conserved throughout the evolutionary process.

\subsection{Structure Universal Relations}
\label{sec:universal_mass_radius}

There is a number of universal relations that connect various parameters of rotating neutron stars, and they are independent to the neutron star's EoS. For example there are equations relating the gravitational binding energy to the compactness \citep{2001ApJ...550..426Lattimer}, the moment of inertia to the compactness \citep{1994ApJ...424..846Ravenhall}, the maximum mass of the non-rotating to the rotating neutron stars \citep{1996ApJ...456..300Lasota,2016MNRAS.459..646Breu}, and the moment of inertia, the quadrupole moment and the Love number, also known as the ``I-Love-Q" relations \citep{2013Yagi,2017PhR...681....1Yagi}. 

In this paper, we focus on universal relations that describe the change in the structure of neutron stars due to rotational effects. To be more specific, we would like to check that the DMANS of this work respect the universal relations of the oblate shape of the rotating neutron stars \citep{2007Morsink,2021PhRvD.103f3038Silva}, and the change of mass and radius of neutron stars due to rotational effects when the central energy density is kept constant \citep{Konstantinou:2022vkr}.

To test the universal relation of the oblate shape of the rotating neutron stars, we compare the polar to the equatorial circumferential radius ratio $R_p/R_e$ to the prediction coming from equation \citep{2021PhRvD.103f3038Silva}
\begin{equation}
R(\theta)=R_{e}\sqrt{\frac{1-e^2}{1-e^2 g(\cos\theta)}},
\label{eq:shape}
\end{equation}
where $e$ and $g(\cos\theta)$ are functions of     $\kappa \equiv \frac{GM}{R_{e}c^2}$, $\sigma \equiv \frac{\Omega^2 R^3_{e}}{GM}$, and empirical coefficients (see Equations (17), (19), and Table 3 in \cite{2021PhRvD.103f3038Silva}).

The mass $M$, and the equatorial radius $R_e$, for a neutron star rotating with a spin frequency $\nu$, can be related to the mass $M_*$ and radius $R_*$ of the corresponding non-rotating NS with the same central density by using the empirical relations \cite{Konstantinou:2022vkr}
\begin{equation}
\begin{aligned}
\frac{R_e}{R_*} =  1 +  \left( e^{A_{2} \Omega_{n2}^2} - 1 + B_{2} \left[ \ln\left( 1 - \left(\frac{\Omega_{n2}}{1.1}\right)^4\right)\right]^2 \right)
\\ \times \left( 1 + \sum_{i=1}^{5} b_{r,i} C_e^i \right),
\end{aligned}
\label{eq:Rinvfit}
\end{equation}
\begin{equation}
\frac{M}{M_*} = 1 + \left( \sum_{i=1}^4 d_i \Omega_{n2}^i\right)  \times \left( \sum_{i=1}^{4} b_{m,i} C_e^i \right),
\label{eq:Minvfit}
\end{equation}
where $C_e = M/R_e$ is the equatorial compactness and $\Omega_{n2}$ is a dimensionless normalized spin frequency dependent on $M$, $R_e$, and $\nu$. The coefficients $A_2$, $B_2$, $b_{r,i}$, $d_i$, and $b_{m,i}$ are listed in Table 1 of \citep{Konstantinou:2022vkr}.
For the DMANS that were created in this work, we use M=$M_{tot}$ and $R_e=R_{eB}$.

The percent deviation of a quantity $Q$ from its corresponding best-fit predicted value $Q_{fit}$
\begin{equation}
 Dev(Q) = 100 \times \frac{Q-Q_{fit}}{Q}.
\end{equation}
is implemented to check if the universal relations are satisfied.

\section{Results and analysis}\label{rs}

\begin{table*}[t!]
\centering
\scriptsize
\caption{Maximum gravitational mass ($M_{\rm max}$), corresponding radius ($R_{\rm max}$), radius at $1.4\,M_\odot$ ($R_{1.4}$), maximum baryonic mass ($M_{\rm b}^{\rm max}$), central temperature at maximum mass ($T_{\rm c}$), moment of inertia at $1.4\,M_\odot$ ($I_{1.4}$), redshift at $M=2.0\,M_\odot$ ($Z_p$), and frequency at maximum mass ($\nu_{\rm max}$) for various EoSs and deformation ratios.}
\label{tab:Mmax_R14_Mb_Tc}
\resizebox{\textwidth}{!}{%
\begin{tabular}{llcccccccccc}
\hline
\textbf{Label} & \textbf{Model} & $r_p/r_e$ & $M_{\rm max}$ [$M_\odot$] & $R_{\rm max}$ [km] & $R_{1.4}$ [km] & $M_{\rm b}^{\rm max}$ [$M_\odot$] & $T_{\rm c}$ [MeV] & $I_{1.4}$ [$10^{45}$ g\,cm$^2$] & $Z_p$ & $\nu_{\rm max}$ [Hz] \\
\hline
\multirow{6}{*}{$s_B=1$, $Y_l=0.4$} 
& No DM & 1.0 & 2.44 & 12.31 & 14.48 & 2.80 & 29.99 & --    & 0.3184 & 0 \\
& No DM & 0.6 & 2.85 & 15.68 & 20.46 & 3.43 & 29.03 & 2.305 & 0.3284 & 1453 \\
& No DM & 0.5 & 2.82 & 17.83 & 24.31 & 3.38 & 29.03 & 2.241 & 0.3290 & 1422 \\
& 5\% DM & 1.0 & 2.25 & 11.80 & 13.83 & 2.51 & 31.56 & --    & 0.3521 & 0 \\
& 5\% DM & 0.6 & 2.58 & 15.20 & 19.61 & 3.02 & 30.24 & 2.059 & 0.3568 & 1449 \\
& 5\% DM & 0.5 & 2.56 & 17.40 & 23.27 & 2.98 & 30.24 & 1.998 & 0.3570 & 1416 \\
\hline
\multirow{6}{*}{$s_B=2$, $Y_l=0.2$} 
& No DM & 1.0 & 2.49 & 12.79 & 15.52 & 2.88 & 71.82 & --    & 0.2922 & 0 \\
& No DM & 0.6 & 2.90 & 16.40 & 22.14 & 3.92 & 69.30 & 2.618 & 0.2987 & 1375 \\
& No DM & 0.5 & 2.87 & 18.71 & 26.36 & 3.83 & 69.30 & 2.538 & 0.2996 & 1343 \\
& 5\% DM & 1.0 & 2.30 & 12.27 & 14.78 & 2.58 & 75.85 & --    & 0.3225 & 0 \\
& 5\% DM & 0.6 & 2.63 & 15.73 & 21.20 & 3.23 & 73.59 & 2.322 & 0.3247 & 1395 \\
& 5\% DM & 0.5 & 2.60 & 18.06 & 25.19 & 3.16 & 73.59 & 2.245 & 0.3259 & 1359 \\
\hline
\multirow{6}{*}{$s_B=2$, $Y_{\nu_e}=0$} 
& No DM & 1.0 & 2.49 & 12.81 & 15.64 & 2.90 & 70.03 & --    & 0.2906 & 0 \\
& No DM & 0.6 & 2.90 & 16.43 & 22.43 & 3.72 & 67.74 & 2.616 & 0.2972 & 1371 \\
& No DM & 0.5 & 2.87 & 18.76 & 26.70 & 3.65 & 67.74 & 2.533 & 0.2981 & 1338 \\
& 5\% DM & 1.0 & 2.24 & 12.16 & 14.82 & 2.54 & 78.35 & --    & 0.3255 & 0 \\
& 5\% DM & 0.6 & 2.53 & 15.98 & 21.37 & 2.98 & 75.10 & 2.291 & 0.3265 & 1349 \\
& 5\% DM & 0.5 & 2.51 & 18.31 & 25.39 & 2.93 & 75.60 & 2.214 & 0.3269 & 1322 \\
\hline
\multirow{6}{*}{$T=0$} 
& No DM & 1.0 & 2.48 & 12.03 & 13.15 & 3.00 & -- & --    & 0.3450 & 0 \\
& No DM & 0.6 & 2.98 & 15.30 & 17.53 & 3.35 & -- & 2.247 & 0.3595 & 1515 \\
& No DM & 0.5 & 2.97 & 17.20 & 20.68 & 3.34 & -- & 2.222 & 0.3603 & 1502 \\
& 5\% DM & 1.0 & 2.22 & 11.53 & 12.71 & 2.61 & -- & --    & 0.3795 & 0 \\
& 5\% DM & 0.6 & 2.60 & 14.81 & 17.28 & 3.09 & -- & 2.006 & 0.3845 & 1496 \\
& 5\% DM & 0.5 & 2.58 & 16.69 & 20.45 & 3.08 & -- & 1.963 & 0.3843 & 1494 \\
\hline
\end{tabular}%
}
\end{table*}

\begin{table*}[t!]
\centering
\scriptsize
\caption{Fixed baryonic masses $M_b=1.55\,M_\odot$ and $M_b=2.30\,M_\odot$. Each cell lists radius $R$ (km) and gravitational mass $M$ ($M_\odot$) at the indicated deformation ratio $r_p/r_e$.}
\resizebox{\textwidth}{!}{%
\begin{tabular}{l|cc|cc|cc|cc|cc|cc}
\hline
& \multicolumn{6}{c|}{$M_b=1.55\,M_\odot$} & \multicolumn{6}{c}{$M_b=2.30\,M_\odot$} \\
\cline{2-13}
& \multicolumn{2}{c|}{$r_p/r_e=1.0$} & \multicolumn{2}{c|}{$r_p/r_e=0.6$} & \multicolumn{2}{c|}{$r_p/r_e=0.5$} & \multicolumn{2}{c|}{$r_p/r_e=1.0$} & \multicolumn{2}{c|}{$r_p/r_e=0.6$} & \multicolumn{2}{c}{$r_p/r_e=0.5$} \\
Model & $R$ & $M$ & $R$ & $M$ & $R$ & $M$ & $R$ & $M$ & $R$ & $M$ & $R$ & $M$ \\
\hline
$s=1,\,Y_l=0.4$ (No DM)      & 14.418 & 1.472 & 20.187 & 1.490 & 23.952 & 1.488 & 13.785 & 2.081 & 18.727 & 2.132 & 21.835 & 2.129 \\
$s=1,\,Y_l=0.4$ (DM 5\%)     & 13.680 & 1.544 & 19.177 & 1.552 & 22.686 & 1.551 & 12.517 & 2.176 & 17.391 & 2.224 & 20.135 & 2.222 \\
\hline
$s=2,\,Y_l=0.2$ (No DM)      & 15.435 & 1.462 & 21.876 & 1.478 & 26.000 & 1.476 & 14.609 & 2.075 & 20.103 & 2.120 & 23.513 & 2.117 \\
$s=2,\,Y_l=0.2$ (DM 5\%)     & 14.595 & 1.534 & 20.726 & 1.539 & 24.560 & 1.539 & 13.328 & 2.170 & 18.635 & 2.211 & 21.663 & 2.208 \\
\hline
$s=2,\,Y_{\nu_e}=0$ (No DM)  & 15.556 & 1.455 & 22.160 & 1.470 & 26.343 & 1.468 & 14.676 & 2.065 & 20.248 & 2.108 & 23.700 & 2.105 \\
$s=2,\,Y_{\nu_e}=0$ (DM 5\%) & 14.619 & 1.528 & 20.864 & 1.533 & 24.722 & 1.533 & 13.078 & 2.167 & 18.611 & 2.200 & 21.610 & 2.199 \\
\hline
$T=0$ (No DM)                & 13.286 & 1.414 & 17.800 & 1.436 & 21.020 & 1.435 & 13.309 & 2.004 & 17.458 & 2.058 & 20.215 & 2.058 \\
$T=0$ (DM 5\%)               & 12.722 & 1.486 & 17.214 & 1.492 & 20.312 & 1.491 & 12.259 & 2.102 & 16.632 & 2.136 & 19.206 & 2.136 \\
\hline
\end{tabular}
}
\label{tab:MgR_fixedMb_all}
\end{table*}

\begin{figure}[!t]	 		
  \includegraphics[width=0.5\textwidth]{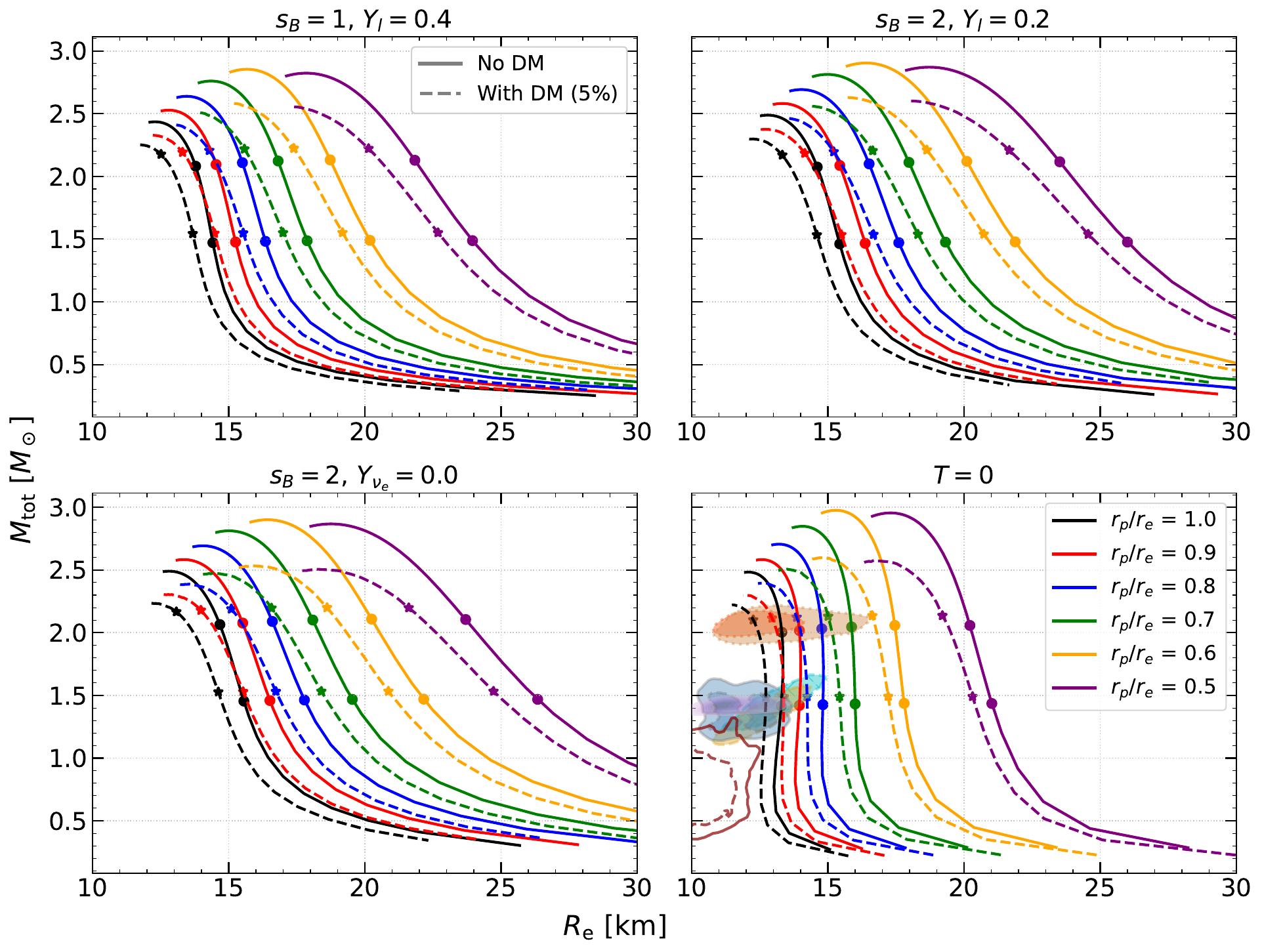}
			 			\caption{The plot shows the relationship between the equatorial radius $R_e$ and the total mass, $M_{\rm tot}$ of the DMANS at different stages of PNS evolution, up to the final stage when the star becomes cold and catalyzed. For the $T=0$ panel, the steel blue area indicates the constraints obtained from the binary components of GW170817, with their respective 90\% and 50\% credible intervals. Additionally, the plot includes the 1 $\sigma$ (68\%) CI for the 2D mass-radius posterior distributions of the millisecond pulsars PSR J0030 + 0451 (in cyan and yellow color) \cite{riley2019, Miller:2019cac} and PSR J0740 + 6620 (in orange and peru color)\cite{riley2021, Miller:2021qha}, based on NICER X-ray observations. Furthermore, we display the latest NICER measurements for the mass and radius of PSR J0437-4715 \cite{Choudhury:2024xbk} (lilac color). The supernova remnant HESS J1731$-$347 \cite{2022NatAs...6.1444D} is shown in red, with the outer contour representing the 90\% CL and the inner contour representing the 50\% CL. The points mark stars with fixed baryon masses of $M_b = 1.55\,M_\odot$ (lower) and $M_b = 2.30\,M_\odot$ (upper); star symbols denote DM-admixed models, while bullet points indicate no-DM cases.
                        }
		\label{Mass_Re}	 	
     \end{figure}

{In \cref{Mass_Re}, the solid lines represent stellar configurations without DM, while the dashed lines correspond to PNSs admixed with mirror DM. From the figure, we present the effect of rotation on the mass-radius diagram of the DMANS at various stages of stellar evolution, with the rotation controlled by the ratio $r_p / r_e$. Generally, the mass and the radius increase as the rotation increases by decreasing $r_p / r_e$, as expected \cite{Konstantinou:2024ynd, Konstantinou:2022vkr}. However, as the DM mass fraction $f_\chi$ increases, the star becomes more compact, leading to simultaneous reductions in both mass and radius. This occurs because the presence of a DM-core adds to the degrees of freedom and total energy density of the stellar matter, effectively increasing the gravitational pull within the star. To maintain hydrostatic equilibrium under this stronger gravitational attraction, the star must contract, raising its central pressure to counterbalance gravity’s effects \cite{Issifu:2025qqw, Kain:2021hpk, Rutherford:2022xeb}. The upper panels depict the neutrino-trapped phase, during which the stellar matter is hot, leading to stars with relatively larger radii. In contrast, the final stage (bottom right panel) shows the star in a cold, catalyzed state, resulting in a more compact structure. The third stage (bottom left panel) represents the point when all neutrinos have escaped the stellar core, and the stellar matter reaches its peak temperature before beginning to cool \cite{Issifu:2024htq}. At this stage, the star remains hot and retains a larger radius compared to its final cold configuration. The bottom panels represent the neutrino-transparent phase of the star's evolution.

In unconstrained sequences (varying central density), the added gravitational potential of DM reduces the baryonic pressure support, shifting the turning point to lower baryon masses and thus decreasing the maximum baryon mass (see \cref{tab:Mmax_R14_Mb_Tc}). In contrast, for fixed baryon-mass configurations, the inclusion of $f_\chi$ increases the total gravitational mass $M_{\rm tot}$ relative to the no-DM case (see \cref{tab:MgR_fixedMb_all} and the markers in \cref{Mass_Re}), as the positive rest-mass contribution from DM outweighs the increase in binding energy (negative), $M_{tot} = M_{\rm GM}^{\rm rest} + M_{\rm DM}^{\rm rest} + E_{\rm bind}^{\rm total}/c^2$. The baryonic fluid becomes more compact and tightly bound, yielding larger $M_{\rm tot}$ and smaller $R$ in the presence of DM. Conversely, along unconstrained sequences, both $M_{\rm tot}$ and $R$ decrease as the star contracts. These trends follow naturally from the additive relation between $M_{GB}$ and $M_{GD}$, where the balance between DM rest mass and baryonic binding energy depends sensitively on the DM distribution and EoS.

Comparing the maximum stellar masses across the panels, we observe that the influence of rotation on PNSs becomes increasingly significant as the star undergoes deleptonization, ultimately allowing a higher maximum mass once it reaches the cold, catalyzed stage. As the PNSs evolve and neutrinos gradually diffuse out, the loss of neutrino pressure causes the star to contract, raising the central density and making the star more compact. This increased compactness enhances rotational stability, enabling the star to spin faster without exceeding the mass-shedding limit. The resulting centrifugal support helps counteract gravity, allowing the star to sustain a higher maximum mass before collapsing. Consequently, the maximum mass is highest in the cold, catalyzed stage, where rotational effects can fully contribute to stabilizing the star’s structure. Moreover, the impact of DM on the star becomes particularly pronounced during the deleptonization phases as well, as the diminishing neutrino pressure reduces the star's support against gravitational collapse, making the star more vulnerable to the effects of DM.

Since direct observational data on the masses and radii of PNSs are unavailable, we constrain the EoS by ensuring that, at the final stage of stellar evolution, the resulting cold, catalyzed NSs are consistent with observed pulsar properties. Accordingly, we include on our plots the confidence contours from pulsars whose masses and radii have been precisely measured by the NICER observatory: PSR J0030+0451~\cite{riley2019, Miller:2019cac} and PSR J0740+60.62~\cite{riley2021, Miller:2021qha}, as well as the inferred mass constraints from NS binaries involved in gravitational-wave events, GW170817 \cite{LIGOScientific:2018cki, LIGOScientific:2017vwq}.
}

\begin{figure*}[t!]
    \centering
    \begin{minipage}[t]{0.9\textwidth}
        \centering
        \includegraphics[width=\textwidth]{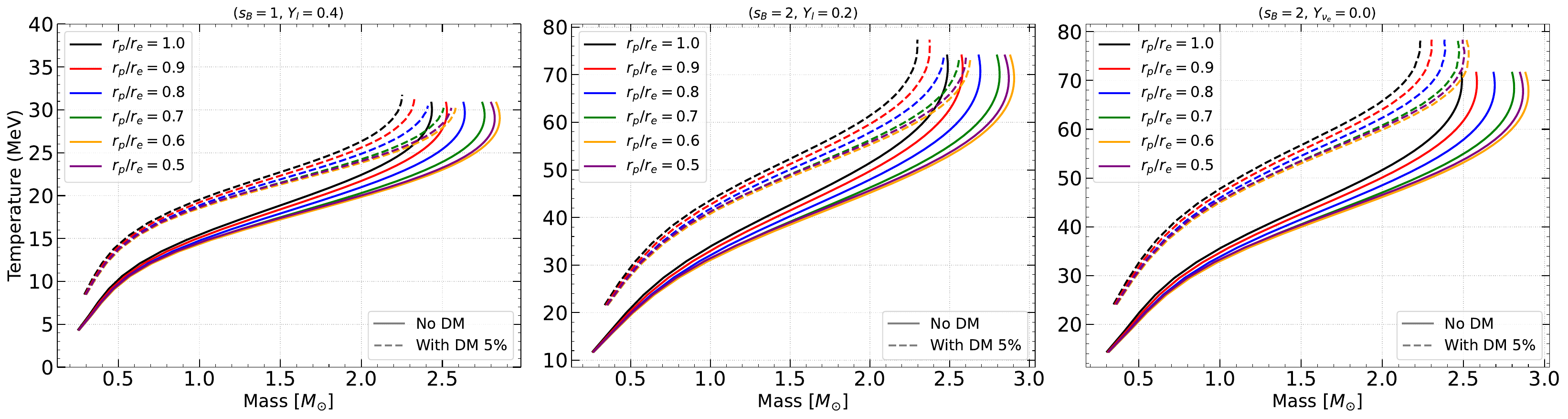}
    \end{minipage}
    \caption{The plots show central temperatures for isentropic models as a function of total mass for different evolutionary stages of rotating PNS admixed with dark matter. }
    \label{fig:temp}
\end{figure*}

In \cref{fig:temp}, we observe that the magnitude of the central temperature behavior in the first stage (first panel), when the star is still trapping neutrinos, is lower than in the second and third stages during the deleptonization process. This behavior is attributed to the higher $Y_l$ and lower $s_B$ at this initial stage \cite{Burrows:1986}. As the star undergoes deleptonization, $Y_l$ decreases while $s_B$ increases, leading to an enhancement of the central temperature behavior. The temperature reaches its peak once all neutrinos have escaped from the stellar interior, after which it begins to decline \cite{Reddy:1997yr}. In each stellar configuration during the evolution, the maximum temperature occurs at the core, which has been computed and recorded in \cref{tab:Mmax_R14_Mb_Tc} as $T_c$ in the unit of MeV.

Although the DMANS, in \cref{fig:temp} correspond to lower maximum masses, they exhibit higher temperatures. This occurs because the introduction of DM into the star increases its compactness, leading to a redistribution of particles within the stellar interior. As interparticle distances decrease, the gravitational potential energy becomes more negative, and according to the virial theorem \cite{Shapiro:1983du}, this requires a corresponding increase in kinetic energy, which manifests as heat \cite{Issifu:2024htq}. Therefore, the presence of mirror DM alters the star’s structure and thermal balance, modifying its thermal evolution and potentially affecting its estimated age. This behavior can be explored by studying deviations in conventional cooling curves, providing an indirect method to probe DM in PNSs during their evolution \cite{deLavallaz:2010wp, Issifu:2025qqw}. Such analyses can potentially lead to physical constraints on the mass and coupling strength of DM particles, offering valuable insights for direct DM searches in terrestrial laboratories \cite{Kouvaris:2011fi}.

From \cref{fig:temp}, another important feature is how rotation leads to a decrease in the central temperature behavior \cite{daSilva:2025cfe}. Across all panels, we observe that a lower ratio $r_p/r_e$ corresponds to lower temperatures, indicating that as the rotation frequency increases, and the stellar mass and radius grow, the temperature in the stellar interior decreases. This occurs primarily because rotation reduces the star’s compactness: centrifugal forces redistribute matter outward, lowering the effective gravitational potential, central density, and pressure. To maintain hydrostatic equilibrium \cite{Stergioulas:2003yp, Burrows:1986} under these conditions, the core temperature must also decrease. Additionally, rotation flattens the entropy gradient \cite{Pons:1998mm, Kastaun:2016yaf}, resulting in a more uniform spread of thermal energy throughout the star rather than concentrating heat in the core. This behavior occurs consistently in stellar configurations both without DM components and in those composed of BM admixed with DM.

\begin{figure}[!t]		 		
  \includegraphics[width=0.5\textwidth]{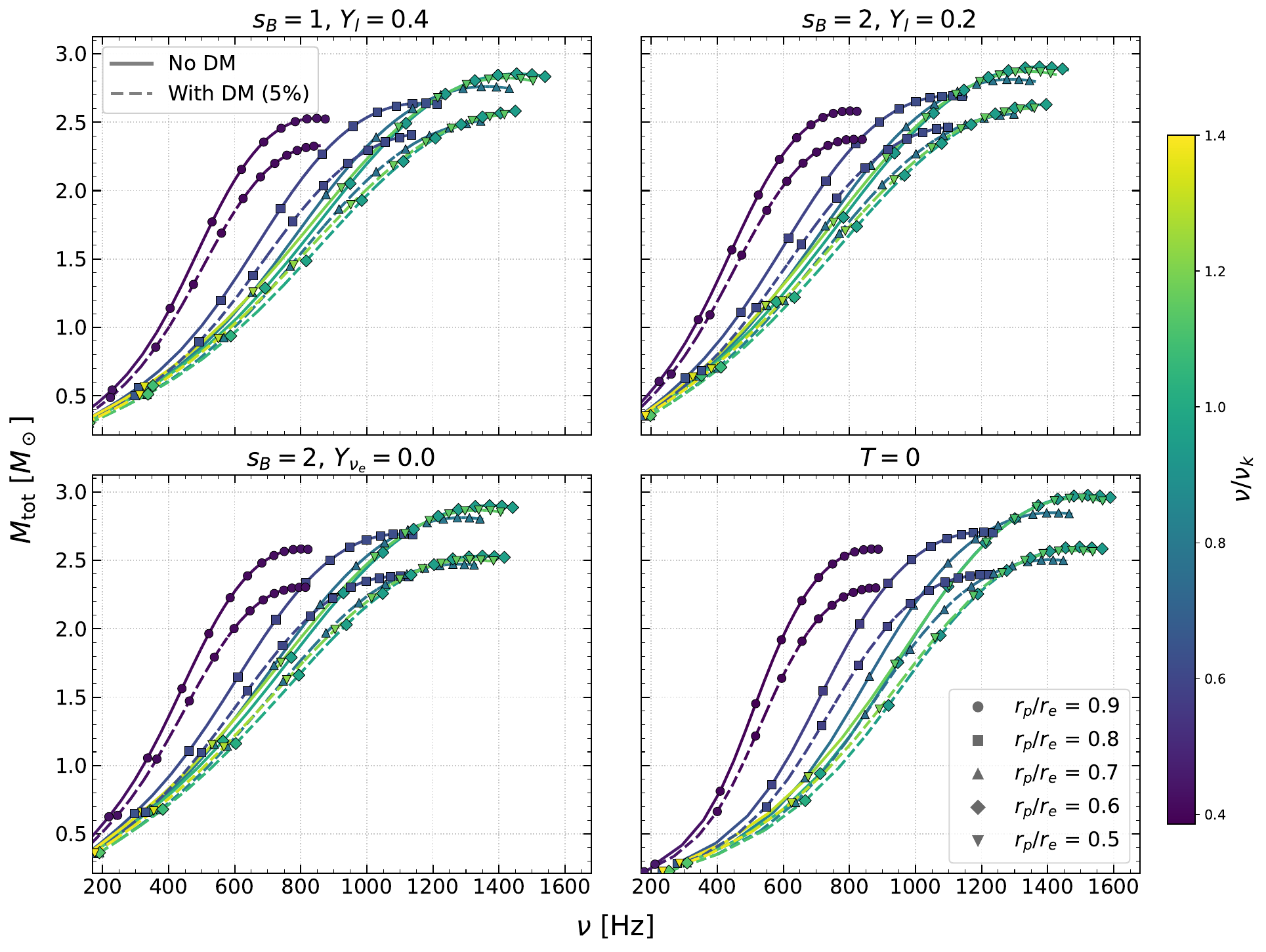}
			 			\caption{Gravitational mass ($M_\mathrm{tot}$) as a function of rotational frequency ($\nu$) for neutron stars under different thermal and lepton conditions: $s_B = 1,\ Y_{l} = 0.4$ (top left), $s_B = 2,\ Y_{l} = 0.2$ (top right), $s_B = 2,\ Y_{\nu_e} = 0$ (bottom left), and $T=0$ (bottom right). Solid lines denote stars without dark matter, while dashed lines include a 5\% dark matter core. Symbols represent varying rotational deformation characterized by polar-to-equatorial radius ratios $r_p/r_e \in \{0.9, 0.8, 0.7, 0.6, 0.5\}$. The color gradient encodes the ratio $\Omega/\Omega_K$.}
		\label{mass_omega}	 	
     \end{figure}

{\Cref{mass_omega} displays the stellar rotation frequency $\nu$ as a function of the total mass $M_{\rm tot}$ for various stages of PNS evolution. The accompanying color bar represents the ratio of the rotation frequency to the Keplerian (mass-shedding) limit, $\nu/\nu_K$ \cite{Stergioulas:1994ea, Cipolletta:2015nga}, detailed derivation of $\nu_K$ and the mass-radius relation can be found in \cite{Glendenning1997}. These plots highlight the enhancement in the maximum supported mass due to rapid rotation, with more massive configurations achievable at higher rotational frequencies. Regions approaching $\nu/\nu_K \approx 1$ correspond to stars nearing the mass-shedding limit, while regions where $\nu/\nu_K \geq 1$ are unphysical, as they exceed the breakup threshold. The mass sequences shown in the plot follow the same trends discussed in \cref{Mass_Re}, where the maximum mass of rotating PNSs increases during deleptonization and reaches its peak when the star becomes cold and catalyzed. This behavior results from the loss of thermal pressure and the increasing compactness as the star evolves. Generally, a decrease in the ratio $r_p/r_e$ indicates greater stellar deformation; under these conditions, the star can sustain higher rotational $\nu$ and support a larger mass, as illustrated in the plots. In our formalism, $r_p/r_e = 1$ corresponds to a spherical star, while $r_p/r_e = 0.5$ signifies a highly oblate configuration \cite{Cook:1993qr}.

As indicated by the color coding, lower-mass stars formed under softer EoS are more susceptible to early mass shedding due to their reduced pressure support. The presence of DM further increases the star’s compactness, modifying the rotationally supported mass-shedding limit. This feature is essential for indirectly probing the presence of DM in NSs through pulsar timing or gravitational-wave observations. Moreover, these structural and dynamical properties are crucial for determining the star’s shape, moment of inertia, quadrupole moments, and spin-down behavior, all of which significantly influence gravitational-wave signatures and pulsar observations. Overall, this plot provides a multi-dimensional probe of NS physics, demonstrating how mass, rotation, thermal state, lepton content, and the presence of a DM core interact. It serves as a valuable theoretical testbed for nuclear physics under extreme conditions, highlighting the stability and evolution of rotating NSs in the presence of DM.

}

\begin{figure}[t!]		 		
  \includegraphics[width=0.5\textwidth]{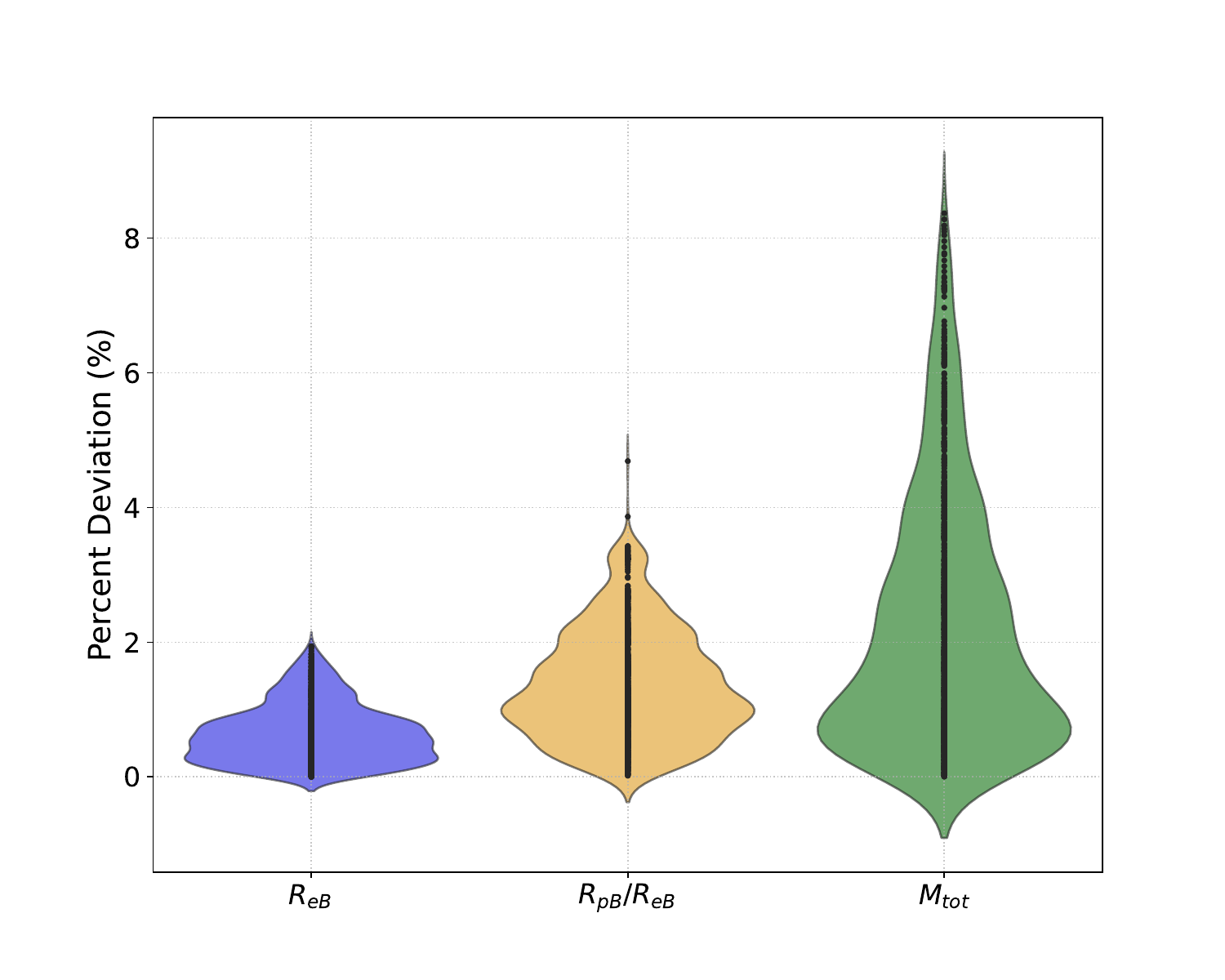}
	\caption{The absolute percent deviation, $|Dev(Q)|$, of the equatorial radius, $R_{eB}$, the polar to equatorial ratio, $R_{pB}/R_{eB}$, and total mass, $M_{tot}$ are shown in violin plots. The data that were produced by the various EoS in this work are visualized as black dots. The curved shape of the violin plot visualizes the data distribution. }
		\label{Univ}	 	
     \end{figure}

Figure \ref{Univ} shows the percent deviations of all of our produced DMANS from the shape universal relations discussed in Section \ref{sec:universal_mass_radius}. All the DMANS that were used here are characterized by a frequency smaller than their corresponding Kepler frequency. We find that $|Dev(R_e)| \lessapprox 1.94 \%$, $|Dev(M_{tot})| \lessapprox 4.69 \%$ and $|Dev(R_{ratioB})| \lessapprox 8.37 \%$. By looking at the data distribution from the violin plots, we see that most of the data are characterized by a percent deviation $\lessapprox 2 \%$, so we can conclude that universal relations are satisfied.

\begin{figure}[!t]		 		
  \includegraphics[width=0.5\textwidth]{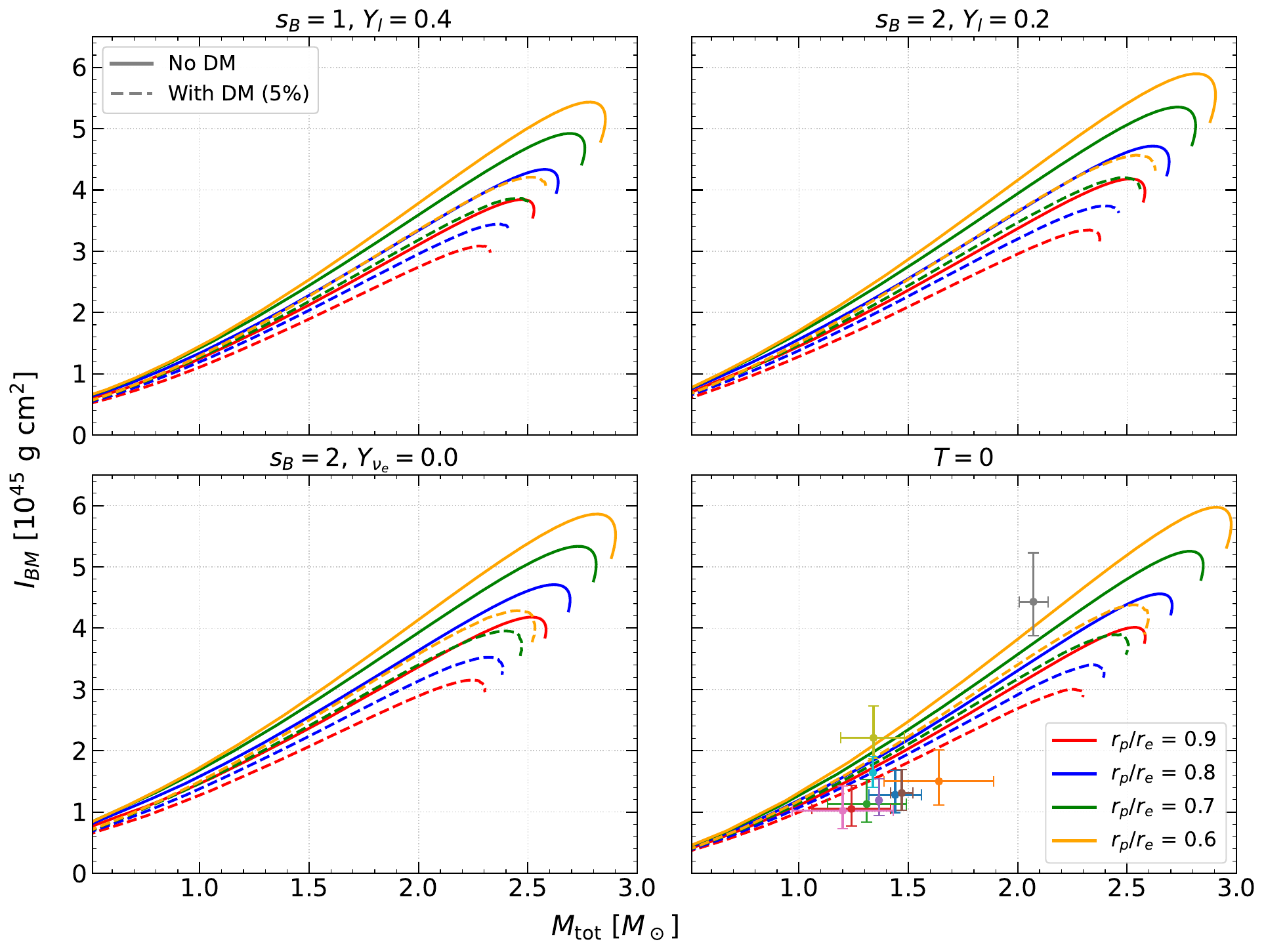}
			 			\caption{ Baryonic moment of inertia ($I_{\rm BM}$) as a function of gravitational mass ($M_{\rm tot}$) for rotating neutron star configurations under different thermal and lepton conditions. Panels correspond to $s_B = 1,\ Y_{l} = 0.4$ (top left), $s_B = 2,\ Y_{l} = 0.2$ (top right), $s_B = 2,\ Y_{\nu_e} = 0$ (bottom left), and $T=0$ (bottom right). Solid lines represent stars without DM, while dashed lines correspond to configurations with 5\% dark matter content. Each color denotes a different rotation rate, defined by the polar-to-equatorial radius ratio $r_p/r_e = \{0.9,\,0.8,\,0.7,\,0.6\}$. Overlaid error bars represent observational constraints from the following pulsars:  
Red: J0437$-$4715,  
Blue: J0751+1807,  
Green: J1713+0747,  
Orange: J1802$-$2124,  
Purple: J1807$-$2500B,  
Brown: J1909$-$3744,  
Pink: J2222$-$0137,  
Gray: J0740+6620 (NICER),  \cite{Li_2022}
Olive: J0030+0451 (NICER), \cite{Silva:2020acr} 
Cyan: J0737$-$3039A (Double Pulsar). \cite{Kumar:2019xgp,PhysRevD.105.063023,Bejger:2005jy}.}
		\label{Mass_IB}	 	
     \end{figure}

{\Cref{Mass_IB} shows the {baryonic} moment of inertia, $I_{BM}$, as a function of the total mass of the PNS evolution to the formation of a cold and catalyzed NS. This plot provides critical insight into the star’s internal structure, composition, and rotational dynamics under varying conditions of $s_B$, $Y_l$, and DM content. $I_{BM}$ primarily quantifies how mass is distributed within the star  \cite{Glendenning1997}. Comparing the curves across the evolutionary sequences, we observe that stars without DM content exhibit larger values of $I_{BM}$ than their DM-admixed counterparts. This indicates that less compact stars are associated with higher moments of inertia. In other words, the stiffer the EoS, i.e., the more resistant the star is to compression, the larger its radius and the higher its $I_{BM}$. Furthermore, examining the panels, we find that as the ratio $r_p/r_e$ decreases, indicating greater stellar deformation due to rotation, $I_{BM}$ increases. This behavior arises from centrifugal flattening, which redistributes mass further from the rotation axis and enhances the moment of inertia.

Additionally, we assess the thermal and lepton effects at different stages of stellar evolution, using the cold, catalyzed stellar configuration as a baseline. We observe that the $I_{BM}$ increases during the second ($Y_l = 0.2,\ s_B = 2$) and third ($Y_{\nu_e} = 0,\ s_B = 2$) stages, when the star is relatively hot and has an expanded radius, compared to the first stage ($Y_l = 0.4,\ s_B = 1$), where the temperature is lower due to the higher lepton content and the radius relatively smaller. The lowest values of $I_{BM}$ occur when the star becomes cold and catalyzed, as evident in \cref{tab:Mmax_R14_Mb_Tc} measured at $1.4\,\rm M_\odot$ \cite{Silva:2020acr}. 
This behavior arises because the EoS softens when thermal pressure causes the star to expand, leading to an increase in the stellar radius and, consequently, a higher moment of inertia. As discussed earlier, the presence of DM does not contribute directly to pressure support but adds to the star’s gravitational mass, which leads to a more centralized distribution of baryons and a reduction in $I_{BM}$. This effect offers a potential avenue for indirectly inferring the presence of DM, since future precise measurements of $I_{BM}$ that deviate from predictions assuming no DM could signal its influence. Thus, for a fixed $M_{\rm tot}$, the $I_{BM}$ for DMANS is higher than that of no DM content. The relationship between $I_{BM}(M_{\rm tot})$ serves as a valuable diagnostic tool for testing the nuclear matter EoS under extreme conditions, probing the presence of DM during PNS evolution, and linking theoretical models to pulsar observations.

}

\begin{figure}[t!]
    \centering
    \begin{minipage}[t]{0.48\textwidth}
        \centering
        \includegraphics[width=\textwidth]{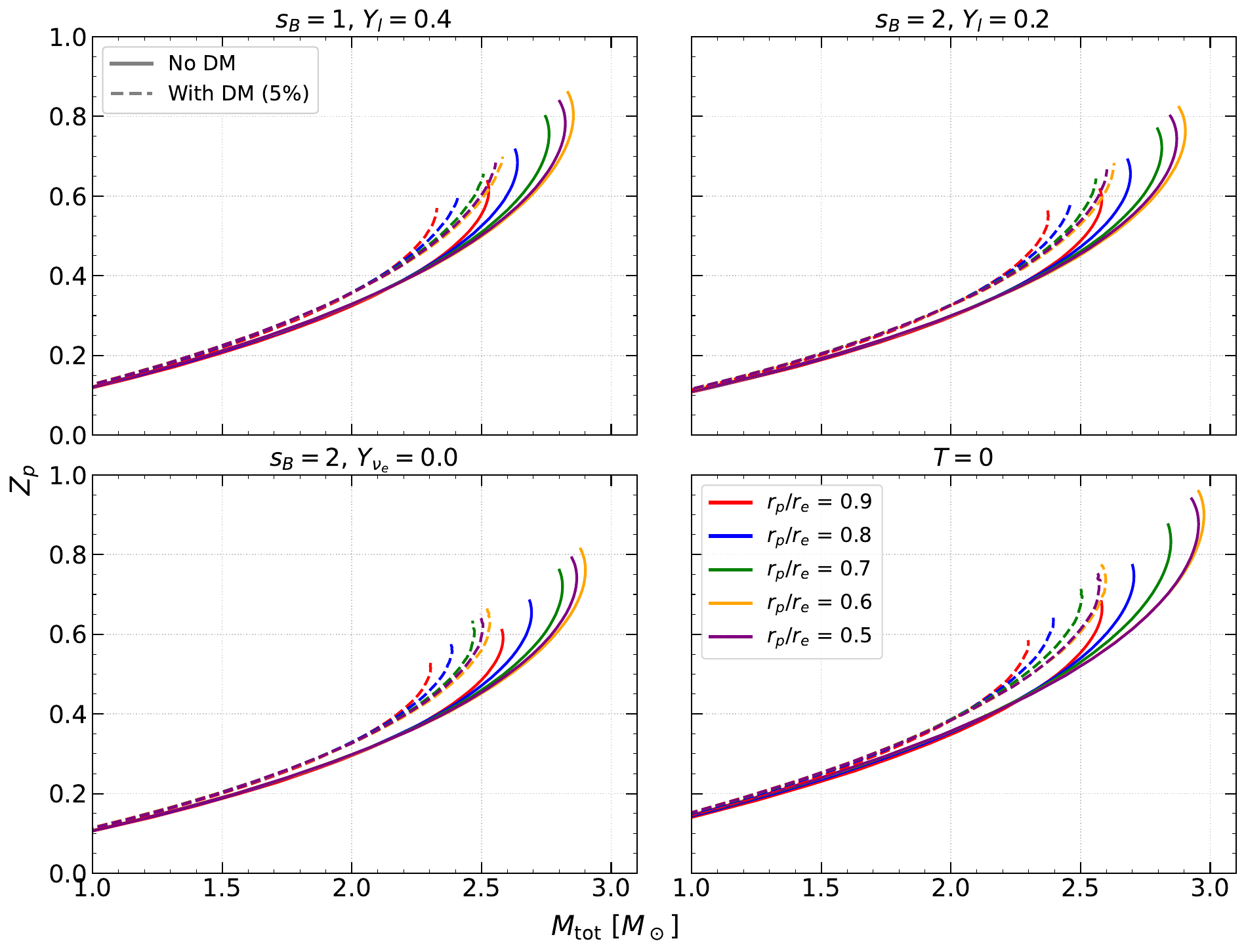}
    \end{minipage}
    \caption{Polar redshift ($Z_p$) as a function of gravitational mass ($M_\mathrm{tot}$) for rotating NS configurations under various thermal and lepton conditions. Each panel corresponds to a different thermodynamic scenario: $s = 1,\ Y_{L,e} = 0.4$ (top left), $s = 2,\ Y_{L,e} = 0.2$ (top right), $s = 2,\ Y_{\nu,e} = 0$ (bottom left), and $T=0$ (bottom right). Solid lines represent stars without dark matter, while dashed lines show configurations with 5\% dark matter content.}
    \label{Mass_Zp}
\end{figure}

{In \Cref{Mass_Zp} we examine polar redshift $Z_p$ as a function of $M_{\rm tot}$.
The plot offers insight into how thermodynamic conditions, such as the $s_B$ and the $Y_l$, together with the presence of DM, influence spacetime curvature, internal stellar structure, and the observable signatures of evolving PNSs. Generally, higher values of the gravitational redshift $Z_p$ indicate stronger gravity and a more compact star, while lower values correspond to weaker gravity and a less compact configuration \cite{1994ApJ...423L.117C, Lattimer:2000nx,Rather:2021yxo}. Comparing the plots at different stages of the stellar evolution, we observe that for a fixed $M_{\rm tot}$, the $Z_p$ decreases for hotter stars, i.e., the second and third stages, compared to the first due to higher radii and increases again when it is cold and catalyzed due to increased compactness. 

Also, because the presence of DM softens the EoS and increases the star’s compactness, it leads to an increase in the $Z_p$ \cite{Leung:2022vcx}. Consequently, along the evolutionary tracks shown in the panels, the dashed curves representing DM-admixed configurations display higher $Z_p$ values than the solid curves corresponding to stars composed purely of OM. Systematic deviations of $Z_p$ from predictions based on visible-matter-only EoS could provide indirect evidence for the existence of a DM core in NSs. Moreover, more massive stars generally exhibit higher $Z_p$ due to their stronger gravitational fields. However, centrifugal forces in rapidly rotating stars cause the star to become oblate, decreasing the $r_p$ and consequently reducing the polar redshift $Z_p$ \cite{1967ApJ...150.1005H, Cook:1993qr}, as is evident when comparing the maximum values of $Z_p$ for the 0.5 and 0.6 cases in the plot. Therefore, $Z_p$ reaches its minimum near the Keplerian limit, where the star is maximally deformed by rotation. Finally, since gravitational-wave emission from NS binaries depends sensitively on stellar compactness and gravitational potential, and X-ray pulsar modeling (e.g., by NICER) constrains the mass and radius, such observations can indirectly probe $Z_p$. This makes measurements of $Z_p$ a valuable potential testbed for investigating the presence of DM in NSs. 

Evidence for gravitationally redshifted absorption lines in NS bursts was first robustly measured by \citet{Cottam:2002cu}, yielding $Z_p = 0.35 \pm 0.05$. Later, pulse-profile modeling by \citet{Miller:2019cac} inferred $Z_p \approx 0.20\text{--}0.25$ for PSR J0030+0451, while \citet{Miller:2021qha} estimated $Z_p \approx 0.27\text{--}0.32$ for the more massive PSR J0740+6620. In the model framework, these estimates coincide with stars with gravitational masses between $1.4 \leq M[M_\odot] \leq 2$ along the evolutionary lines.

}

\section{Final remarks and conclusion} \label{conc}
{Our investigation explores the global evolution of rotating NSs admixed with mirror DM, examining how rotation and DM jointly influence stellar properties from the early neutrino-trapped phase to the mature, neutrino-transparent, cold, and catalyzed configurations. The analysis reveals competing physical mechanisms: DM increases stellar compactness through stronger gravitational binding, reducing both mass and radius \cite{Leung:2022wcf, Kain:2021hpk}, while rotational support counteracts gravity, increasing these parameters up to the Keplerian limit where stellar deformation becomes maximal \cite{Glendenning1997}. Notably, we find that rotational stability is reduced in the hotter intermediate evolutionary stages (the second and third phases) compared to the cooler initial and final stages, with the presence of DM further amplifying these instabilities.

The thermal behavior exhibits contrasting trends: rotation lowers the core temperature by redistributing thermal energy over a larger volume and reducing central density, whereas DM raises the temperature by increasing stellar compactness, deepening the gravitational potential, and enhancing thermal energy in accordance with the virial theorem \cite{Issifu:2024htq}. This effect is distinctive to DM. By contrast, the inclusion of strange particles and other exotic degrees of freedom softens the EoS but redistributes the thermal energy among a greater number of species, thereby lowering the temperature \cite{daSilva:2025cfe, Issifu:2024fuw}. These competing processes, centrifugal expansion versus DM-induced contraction and heating, produce distinctive observational signatures in thermal emission and structural properties, potentially offering indirect probes of DM within NSs. The main findings are summarized in \cref{tab:Mmax_R14_Mb_Tc}.

 Our systematic comparison between the PNS EoS and cold NS universal relations \cite{Konstantinou:2022vkr, 2021PhRvD.103f3038Silva} reveals striking agreement, with deviations limited to $\leq$5\% in $M_{\text{tot}}$, $\leq$2\% in $R_e$, and $\leq$8\% in the flattening ratio ($R_p/R_e$), all consistent with theoretical expectations for hot, lepton-rich matter. The $I_{\text{BM}}$ emerges as a particularly powerful diagnostic: its strong dependence on rotation (increasing $I_{\text{BM}}$) and DM content (decreasing $I_{\text{BM}}$) directly traces mass redistribution, with peak sensitivity during the hot, intermediate phase \cite{daSilva:2025cfe}. This phase-dependent response establishes $I_{\text{BM}}$ as a relevant probe of finite-temperature EoS effects inaccessible in cold NS studies.  

Furthermore, we identify a clear monotonic trend in $Z_p$: it increases systematically with both rotation (due to centrifugal support) and additional DM admittance, driven by the resulting increase in stellar compactness. This dual dependence makes $Z_p$ a promising observable for constraining DM properties in multi-messenger observations \cite{Miller:2019cac, Miller:2021qha}. The exceptional consistency (scatter $\leq$2\%) across all tests demonstrates that universal relations remain robust even for the extreme thermal and compositional gradients in PNSs, providing a critical link between supernova remnants and mature NS populations. These results underscore the need to develop next-generation universal relations that incorporate rotational and thermal corrections, which will be essential for interpreting data from advanced gravitational wave detectors (e.g., the Einstein Telescope, Cosmic Explorer) and X-ray observatories (e.g., STROBE-X). Such refined relations could potentially resolve long-standing uncertainties in the high-density EoS and enable direct detection of exotic components like DM in neutron stars.

}

\begin{acknowledgments}

A.I. acknowledges financial support from the São Paulo State Research Foundation (FAPESP), Grant No. 2023/09545-1. 
This work is part of the project INCT-FNA (Proc. No. 464898/2014-5). T.F. is supported by the National Council for Scientific and Technological Development (CNPq) under Grants Nos. 306834/2022-7. T. F. also thanks the financial support from  Improvement of Higher Education Personnel CAPES (Finance Code 001) and FAPESP Thematic Grants (2023/13749-1 and 2024/17816-8). A.K. acknowledges financial support from "The three-dimensional structure of the nucleon from lattice QCD (3D-N-LQCD)” program, funded by the University of Cyprus. P.~Thakur is supported by the National Research Foundation of Korea (NRF) grant funded by the Korea government (MSIT) (No.~RS-2024-00457037).

\end{acknowledgments}

\bibliography{refrences}
\end{document}